\documentclass[journal=nalefd,manuscript=article]{achemso}

\usepackage[version=3]{mhchem} 
\usepackage{xcolor}

\author{Bingqiang Ji}
\affiliation{Mechanical Science and Engineering, University of Illinois at Urbana-Champaign, Urbana, IL 61801, USA}
\altaffiliation{These two authors contributed equally.}
\author{Amrit Singh}
\affiliation{Mechanical Science and Engineering, University of Illinois at Urbana-Champaign, Urbana, IL 61801, USA}
\altaffiliation{These two authors contributed equally.}
\author{Jie Feng}
\affiliation{Mechanical Science and Engineering, University of Illinois at Urbana-Champaign, Urbana, IL 61801, USA}
\alsoaffiliation{Materials Research Laboratory, University of Illinois at Urbana-Champaign, Urbana, IL 61801, USA}
\email{jiefeng@illinois.edu}

\title[]
  {Water-to-air transfer of nano/micro-sized particulates: enrichment effect in bubble bursting jet drops}


\begin{document}

\begin{tocentry}

\centering
\includegraphics[width=5.7cm]{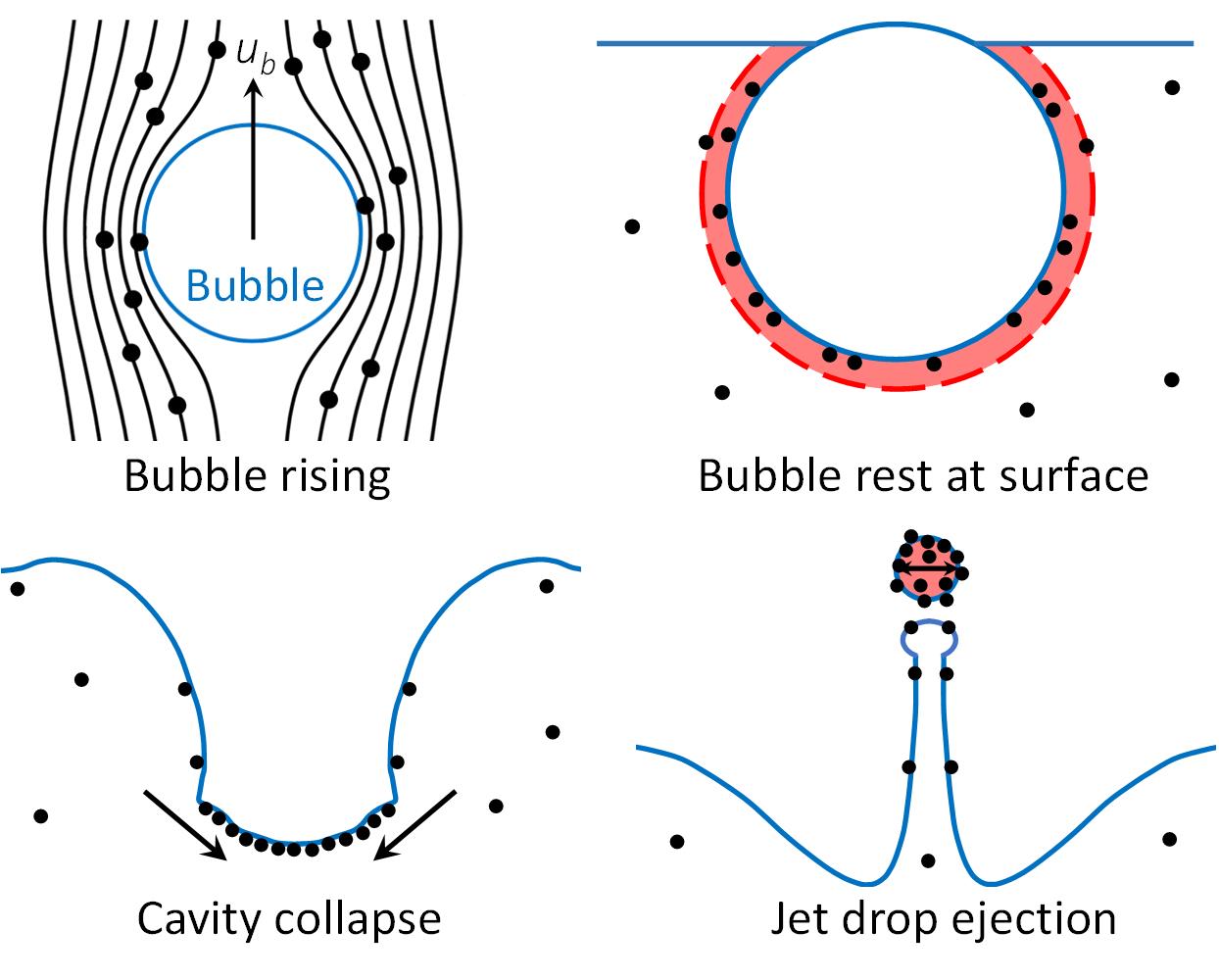}





\end{tocentry}

\begin{abstract}

Bubbles dispersed in liquids are widely present in many natural and industrial processes, and play a key role in mediating mass transfer during their lifetime from formation to rising to bursting. In particular, nano/micro-sized particulates and organisms, present in the bulk water can be highly enriched in the jet drops ejected during bubble bursting, impacting global climate and public health. However, the detailed mechanism of this enrichment remains obscure, with the enrichment factor being difficult to predict. Here, we experimentally investigate the enrichment of nano/micro-sized particles in bubble bursting jet drops and highlight the underlying hydrodynamic mechanism, combining the effects of bubble scavenge and bursting on the transport of particles. Scaling laws for the enrichment factor are subsequently proposed that describe both our and prior experimental results reasonably well. Our study may provide new insights for water-to-air transfer of microbes related to bubble bursting.
\end{abstract}

\section{Introduction}
Bursting of bubbles at liquid surfaces is a ubiquitous process which generates film drops by cap disintegration and jet drops by the fragmentation of the upward liquid jet induced by cavity collapse \cite{knelman1954mechanism, deike2022mass, lewis2004sea, jiang2022submicron}. In natural water bodies, bubbles are typically smaller than a millimeter in diameter \cite{deane2002scale, edzwald1995principles, cartmill1993bubble, blenkinsopp2010bubble}, and mainly produce jet drops with a typical size of several to dozens of microns instead of film drops \cite{brasz2018minimum, blanco2020sea, berny2021statistics, ganan2021physics, veron2015}. These small drops may remain suspended in the air, containing various compositions from the bulk water \cite{wang2017role}, such as sea salts \cite{o1997marine,lewis2004sea}, surfactants \cite{chingin2018enrichment, o2004biogenically}, oil spills \cite{ji2021compound, sampath2019aerosolization}, nanoplastic and microplastic particles \cite{masry2021experimental, allen2020examination}, and even nano/micro-sized organisms \cite{baylor1977water,blanchard1983production, walls2014moving}. Therefore, bubble bursting jets play an important role in mediating mass transfer across the air-water interface in a wide range of industrial, geological, and biological phenomena, including the flavor release from sparkling beverages \citep{Seon2017Effervescence}, sea spray aerosol generation \cite{lewis2004sea, veron2015, spiel1995births}, oceanic vegetative reproduction \cite{hariadi2015determining}, and airborne pathogen transmission \cite{blanchard1970, joung2017bioaerosol}.

It has been widely recognized that both soluble and insoluble components can be substantially enriched in jet drops with a much higher concentration compared to that in the bulk water, up to a thousand fold \cite{chingin2018enrichment, blanchard1989ejection, burrows2014physically, sakai1988enrichment, sakai1989ion}. The enrichment of nano/micro-sized particles is of particular interest considering the transport of microbes by bursting bubbles in contaminated water \cite{walls2014moving, blanchard1970}, which may be one potential mechanism that increases the risk of airborne pathogen transmission \cite{bourouiba2020, 2021LouThe}. This enrichment is believed to be caused by the scavenging of suspended particles by the bubble during rising, and subsequently, the collection of the captured particles along with the liquid shell around the bubble cavity surface into the jet drops upon bursting \cite{macintyre1972flow, sakai1988enrichment, blanchard1989ejection}. A majority of the prior studies investigated bacteria enrichment by collecting the jet drops and then culturing the bacteria for measurement, with different bubble rising distances, bubble sizes, bacteria types and concentrations, as well as culture ages \cite{syzdek1985influence, burger1985droplet, sakai1992fractionation, hejkal1980water, blanchard1970, blanchard1972concentration, blanchard1981bubble, bezdek1972surface}. However, due to the limitations of such an indirect measurement and the complexity of the biological systems, the enrichment factor of the bacteria reported by prior studies varied dramatically from 1 to 1000 even under the similar experimental conditions, and the detailed hydrodynamic mechanism is still elusive \cite{blanchard1978seven}. Therefore, predicting the enrichment factor of particulate matter in bubble bursting jet drops remains a formidable challenge.

Here, we experimentally investigate the enrichment of nano/micro-sized spherical particulates in bubble bursting jet drops in a controlled dispersal system by direct visualization of bubble dynamics and measurement of particulate concentrations. We provide the detailed hydrodynamic mechanism determining particle enrichment, and derive scaling laws which quantify the dependencies of particulate enrichment factor on the bubble and particle sizes, and the bubble rising distance. The proposed scaling laws quantitatively agree with our experimental results over a wide range of experimental parameters. Furthermore, for the first time, the scaling laws describe the main trend of the enrichment of bacteria in previous experiments, potentially advancing the framework for the modeling of water-to-air transmission of particulates involving bubble bursting jets.

\section{Results and discussion}

\subsection*{Particulate enrichment in top jet drops}

\begin{figure*}
\centering
\includegraphics[width=16.5cm]{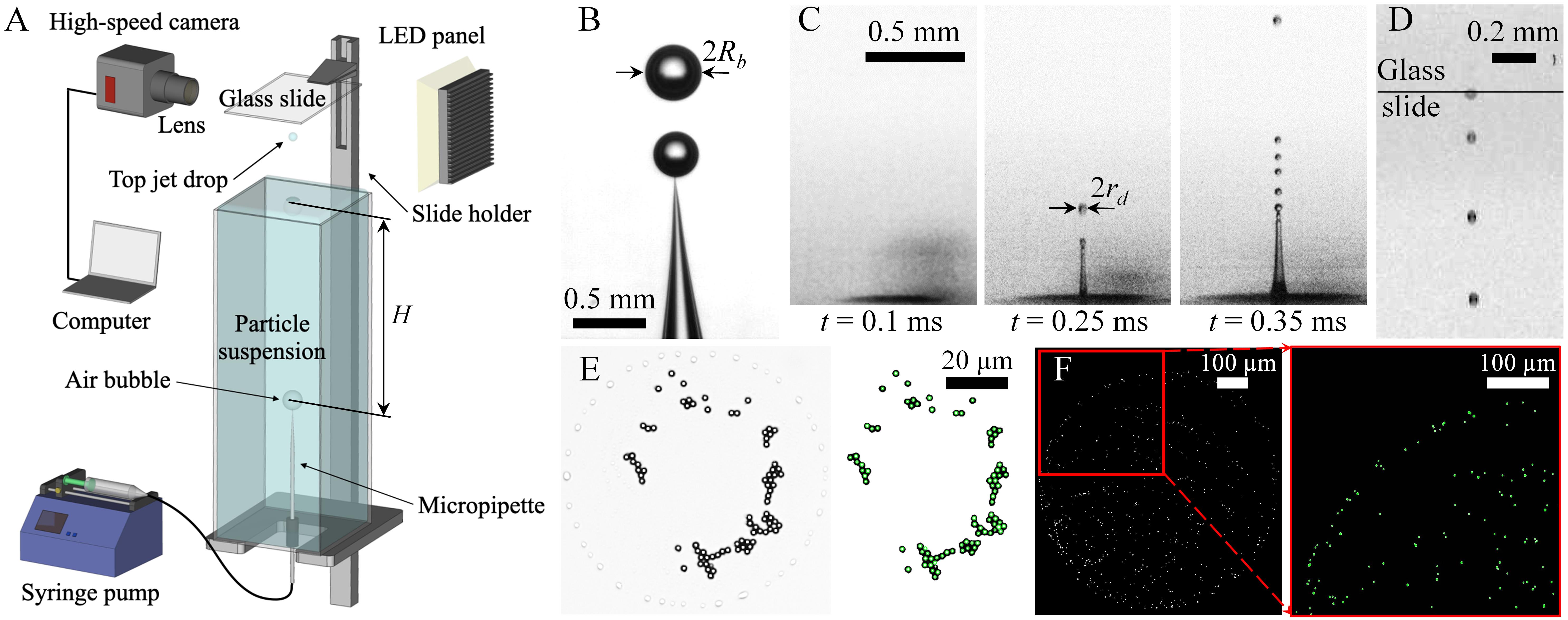}
\caption{(A) Experimental setup. An air bubble is injected using a micropipette at a distance $H$ beneath the surface of the aqueous particle suspension. After rising to the liquid surface, the bubble rests and then bursts, ejecting jet drops upwards. Only the top jet drop is collected by the glass slide. A high-speed camera is used to measure the bubble radius $R_b$ and the top jet drop radius $r_d$. Experimental images show (B) a bubble of $R_b = 0.19$ mm forming and rising ({\color{blue}Movie S1}) before (C) bursting at the liquid surface and ejecting jet drops ({\color{blue}Movie S2}). Finally (D) the top jet drop of $r_d = 21$ $\mu$m is collected by a glass slide ({\color{blue}Movie S3}). The images are superimposed every $\Delta t = 31.25$ ms and 0.2 ms for (B) and (D), respectively. (E) Original (left) and processed images with the particles identified by green dots (right) of the dried top jet drop, with $r_d = 34$ $\mu$m and $r_p = 1$ $\mu$m. (F) Original (left) and processed (right) fluorescent images of a 0.33 $\mu$L drop of bulk particle suspension dried on the glass slide, with $r_p = 1$ $\mu$m.}
\label{fig:experiments}
\end{figure*}

\begin{figure*}
\centering
\includegraphics[width=16.5cm]{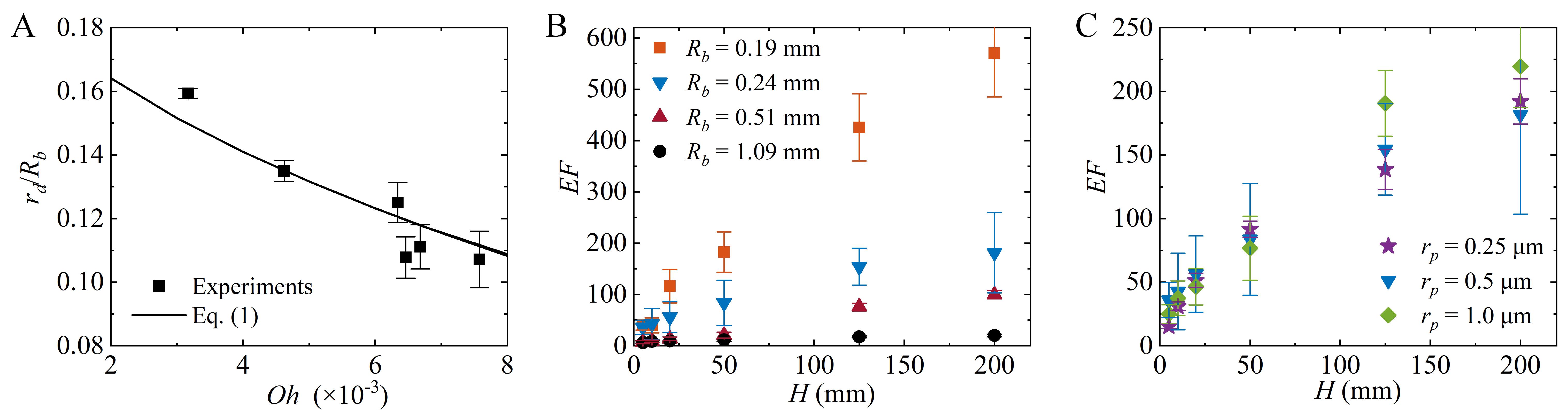}
\caption{(A) Top jet drop radius $r_d$ scaled by the bubble radius $R_b$ as a function of the Ohnesorge number $Oh (= \mu/\sqrt{\rho_wR_b\gamma})$. Experimental results are captured by Eq. \ref{eqn:rd} \cite{blanco2021jets}. Error bars represent the standard deviation of data of at least 3 runs. (B) The enrichment factor $EF$ of particles of radius $r_p = 0.5$ $\mu$m in the top jet drop as a function of bubble rising distance $H$ with $R_b = 0.19-1.09$ mm. $EF$ increases with increasing $H$ and decreasing $R_b$. (C) $EF$ of particles in the top jet drop as a function of bubble rising distance $H$ with $r_p = 0.25-1$ $\mu$m and $R_b = 0.28 \pm 0.04$ mm, showing $EF$ is independent of $r_p$ in the  current experiments. Error bars represent the standard deviation of data of at least 10 runs.}
\label{fig:raw data}
\end{figure*}

In the experiments (Figure \ref{fig:experiments}(A)), an air bubble of radius $R_b$ is generated using a micropipette at a distance $H$ beneath the surface of the aqueous suspension of fluorescent polystyrene (PS) particles with a number concentration of $C_b \approx 2.3 \times 10^6$/mL (Figure \ref{fig:experiments}(B), {\color{blue}Movie S1}). After rising to the water surface, the bubble rests and then bursts, ejecting jet drops upwards (Figure \ref{fig:experiments}(C), {\color{blue}Movie S2}). We focus on the top jet drop ejected by the bursting bubble, which has the highest enrichment as shown by previous studies \cite{blanchard1978seven, sakai1988enrichment, blanchard1970}. The formation and bursting of the bubble are visualized by a high-speed camera to measure $R_b$ and the top jet drop radius $r_d$ (See details in \textit{\color{blue}Methods}). For a bubble bursting at a pure water surface, the top jet drop size is determined only by the Ohnesorge number ($Oh = \mu/\sqrt{\rho_wR_b\gamma}$ with $\mu$, $\rho_w$, and $\gamma$ representing the water viscosity, density, and surface tension, respectively, evaluating the effects of viscosity to inertial and surface tension) when the Bond number ($Bo = \rho_wgR_b^2/\gamma$ with gravitational acceleration $g$, evaluating the effects of gravity relative to surface tension) is negligible ($Bo < 0.16$ in our experiments) \cite{ganan2017revision, walls2015jet, gordillo2019capillary}. Figure \ref{fig:raw data}(A) shows the dimensionless top jet drop radius $r_d/R_b$ observed experimentally as a function of $Oh$, which is well described by the scaling law proposed by a previous study \cite{blanco2021jets} on bubble bursting at a pure water surface,
\begin{equation}
\frac{r_d}{R_b}=0.22(1-(\frac{Oh}{0.031})^{0.5}), \label{eqn:rd}
\end{equation}
indicating that the presence of the particles used in our experiments does not affect the jet dynamics of bubble bursting. In addition, the surface tensions of all the particle suspensions we used were confirmed to remain the same as that of water using the pendent drop method.

Only the top jet drop is collected using a glass slide placed above the water surface, verified by the high-speed camera (Figure \ref{fig:experiments}(D), {\color{blue}Movie S3}). The particle numbers in the jet drop and bulk water, $N_d$ and $N_b$, respectively, are quantified using microscope images of the collected dried jet drop (Figure \ref{fig:experiments}(E)) and a dried bulk drop of volume $V_b$ on a glass slide (Figure \ref{fig:experiments}(F)), by either the bright field or fluorescent light. The enrichment factor ($EF$) of particulates in the top jet drop is calculated as
\begin{equation}
EF=\frac{N_d/V_d}{N_b/V_b}=\frac{3N_d}{4\pi r_d^3C_b}. \label{eqn:EF}
\end{equation}

We use nano/micro-sized spherical particles of $r_p=$ 0.25, 0.5, and 1 $\mu$m as a representative model for viruses and bacteria based on size, with $R_b = 0.19-1.09$ mm and $H = 5- 200$ mm. As shown in Figures \ref{fig:raw data}(B) and \ref{fig:raw data}(C), we observe an $EF$ varied between 6 and 571, which is consistent with that reported by previous studies \cite{sakai1988enrichment, blanchard1989ejection}. Specifically, $EF$ increases with the bubble rising distance $H$ in approximate linearity since more particles may be captured by the rising bubble with a larger $H$. In addition, $EF$ increases significantly with decreasing bubble radius $R_b$, which is approximately 571 and 18 when $R_b = 0.19$ and 1.09 mm with $H = 200$ mm, respectively. Furthermore, we find that the particle radius $r_p$ barely affects $EF$ at a range of $r_p = 0.25-1$ $\mu$m. To rationalize the effects of different parameters on $EF$, we will further discuss the bubble hydrodynamics contributing to the particulate enrichment in the top jet drop, including the scavenging of particles during bubble rising, and the collection of particles into the jet drop upon bursting.

\subsection*{Particle scavenging by bubble rising}

\begin{figure*}
\centering
\includegraphics[width=12.5cm]{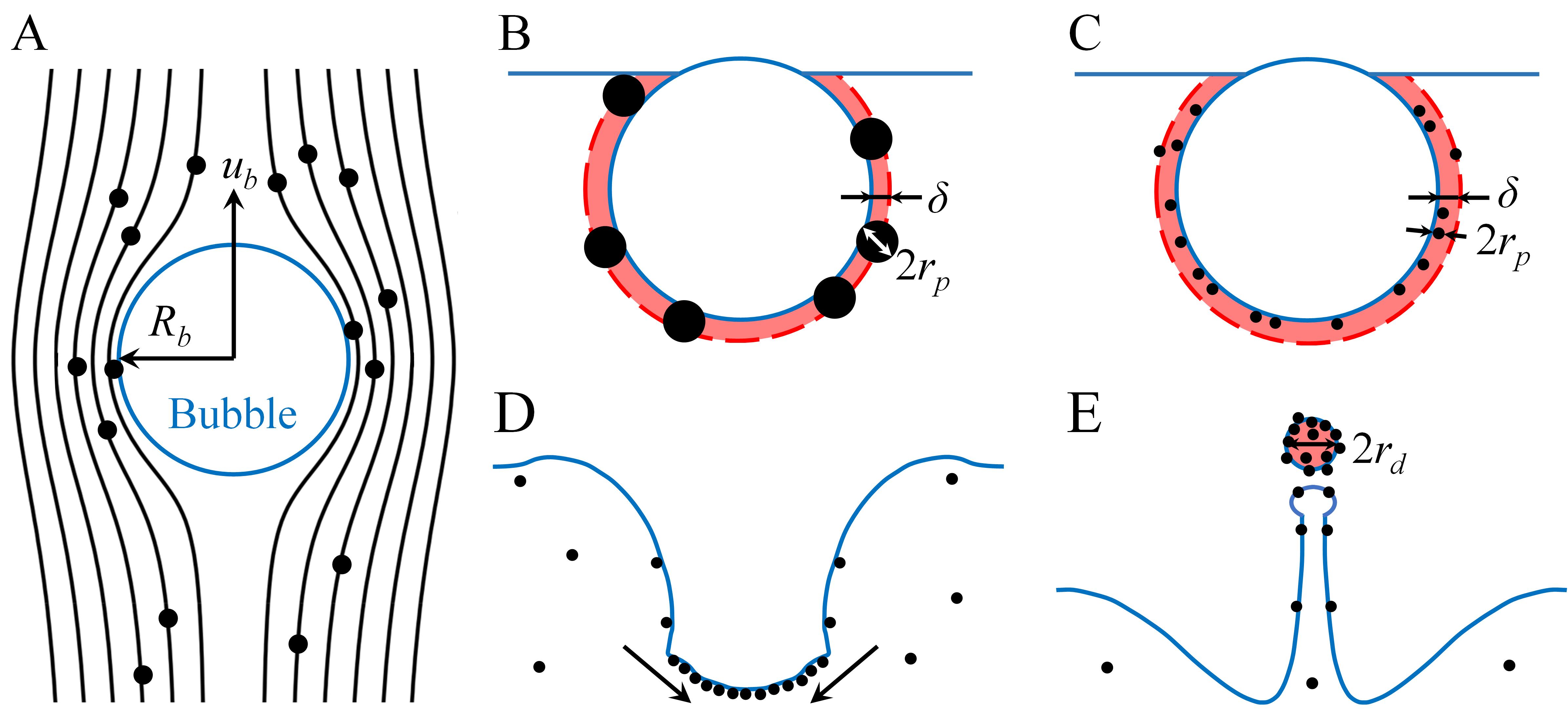}
\caption{Mechanism of particulate enrichment in bubble bursting jet drop. (A) A rising bubble collects particles from the contaminated water by interception. Small particles with negligible inertia follow the streamlines, with only those close to the bubble surface being intercepted and carried by the bubble. Sketches of a bubble carrying particles at the water surface where (B) the particle diameter $2r_p$ is larger than the thickness of the liquid shell $\delta$ forming the top jet drop after bursting (Regime I) or (C) $2r_p < \delta$ (Regime II). (D) During bubble bursting, bubble cavity collapse excites capillary waves propagating along the bubble cavity surface, which sweep the particles towards the bubble bottom pole. (E) A jet then forms and eventually pinches off, ejecting a top jet drop with a enriched particle concentration.}
\label{fig:mechanism}
\end{figure*}

The suspended particles are scavenged by the bubble during rising, as illustrated by Figure \ref{fig:mechanism}(A), where the particle size is much smaller than the bubble radius ($r_p \ll R_b$). Therefore, the concentration of particles at the rising bubble surface will become enriched compared with that of the bulk water. In the current experiments, considering the small $r_p$ and minor density difference between particle and water, gravitational effects can be ignored \cite{weber1983mechanism}. Meanwhile, for particles with $r_p > 0.1$ $\mu$m, diffusive effects can also be neglected \cite{walls2017}. Additionally, the Stokes number (comparing the particle response time to the fluid characteristic time scales), $St = (\rho_p-\rho_w)r_p^2u_b/(9\mu R_b)$ with $\rho_p$ the particle density and $u_b$ the bubble velocity, of the particles in our experiments is $< 10^{-4}$, indicating that the particles can be considered as inertialess, and follow the flow streamlines. Thus, the capture of suspended particles by the rising bubble is dominated by the interception mechanism \cite{walls2017, weber1983mechanism}. Considering the Reynolds number ($Re = 2\rho_wu_bR_b/\mu \approx 20-600$, comparing the inertial to viscous effects) of the rising bubbles in our experiments, the flow around the bubble can be well approximated as potential flow, allowing for the neglect of viscous and rotational effects \cite{manica2016hydrodynamics}. Given such assumptions, a distance from the bubble center-line under which all particles collide and attached can be derived as $\sqrt{3r_p/R_b}$ \cite{sutherland1948physical,walls2014moving}, which gives the collision efficiency of the particles in the liquid volume ($V_s = \pi R_b^2H$) swept by the rising bubble, $E_c = 3r_p/R_b$. In our experiments, all bubbles rise in a rectilinear trajectory without significant path instability. In addition, we assume the small particles are attached to the surface upon collision and the coating does not change the surface mobility significantly. Therefore, the number of particles attached to the rising bubble surface can be calculated as $N_c = E_cV_sC_b = 3\pi R_br_pHC_b$.

\subsection*{Particle collection during bubble bursting}

\begin{figure*}
\centering
\includegraphics[width=14cm]{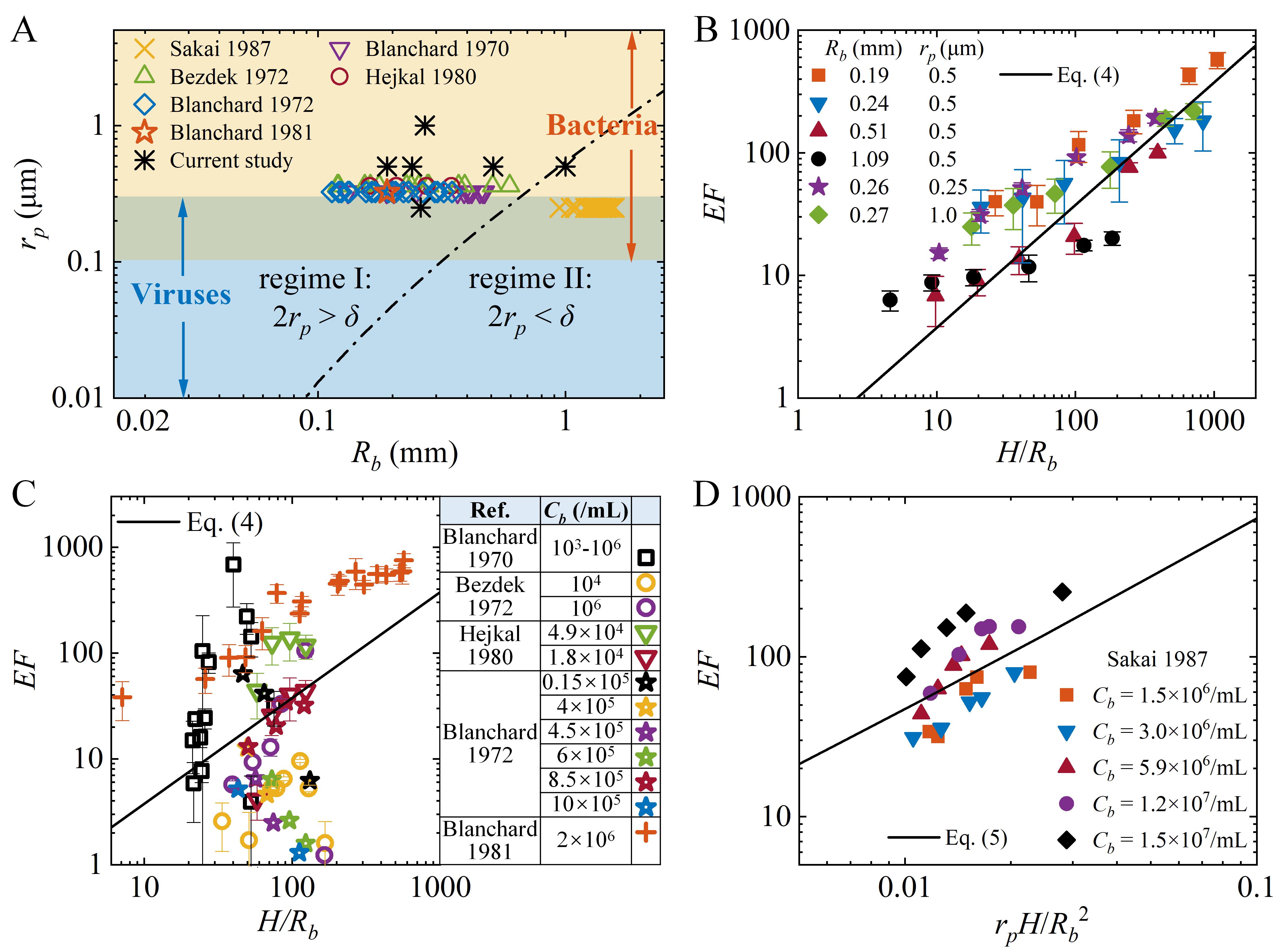}
\caption{(A) Regime map for cases where $2r_p > \delta$ and $2r_p < \delta$ regarding the bubble radius $R_b$ and particle radius $r_p$, with the boundary between the two regimes denoted by the dash-dotted line. The parameter range of previous studies \cite{sakai1988enrichment,blanchard1970,bezdek1972surface,hejkal1980water,blanchard1972concentration,blanchard1981bubble} and this study are shown as scatters. The blue and yellow areas represent the size range of viruses and bacteria, respectively. (B) The enrichment factor $EF$ of particles in the top jet drop as a function of the normalized bubble rising distance $H/R_b$ in regime I. Eq. \ref{eqn:EF 1} (line) agrees with the experimental results (scatters) well over a wide parameter range. (C) Comparison between the model-predicted $EF$ (Eq. \ref{eqn:EF 1}, line) and the experimental results (scatters) in previous studies using rod-shaped (0.25 $\mu$m radius and 1-2 $\mu$m length) and spherical (0.25-1.5 $\mu$m radius) bacteria in regime I, where $R_b$ is calculated using Eq. \ref{eqn:rd} based on the reported $r_d$ \cite{blanchard1970,bezdek1972surface,hejkal1980water,blanchard1972concentration,blanchard1981bubble}. (D) Enrichment factor $EF$ of spherical PS particles in the top jet drop as a function of $r_pH/R_b^2$ in regime II with $r_p = 0.25$ $\mu$m and $H$ = 100 mm. Eq. \ref{eqn:EF 2} (line) predicts the previous experimental results (scatters) \cite{sakai1988enrichment} well with a proportionality coefficient of 8.} 
\label{fig:modelprediction}
\end{figure*}

\noindent
After the bubble reaches the water surface along with the captured particles, it eventually bursts and accumulates the particles at the bubble cavity surface into the ejected drop, further increasing the particle concentration compared to that in the bulk water, as illustrated by Figures \ref{fig:mechanism}(B-E). Due to the sweep of the capillary waves excited by the cavity collapse, the liquid layer beneath the bubble surface is accumulated at the bottom of the bubble during cavity collapse before being ejected upwards, forming jet drops \cite{blanco2021jets,gordillo2019capillary}. Based on this insight, MacIntyre \cite{macintyre1972flow} first proposed an assumption that the top jet drop originates from the liquid shell just beneath the bubble surface (Figures \ref{fig:mechanism}(B-E)). This assumption has been validated by numerical simulations on bubble bursting at surfactant-laden surfaces which illustrated the enrichment of surfactants in the jet drop \cite{constante2021dynamics, boulton1995effect, dey1997experimental}, and has also been successfully adopted in the theoretical model predicting the $EF$ of surface-active compounds \cite{chingin2018enrichment}. By balancing the top jet drop volume ($V_d = 4\pi r_d^3/3$) and the subsurface liquid shell volume ($V_{ls} = 4\pi R_b^2\delta$), the thickness of the subsurface liquid shell ($\delta$) forming the top jet drop is estimated as
\begin{equation}
\delta=0.0036(1-(\frac{Oh}{0.031})^{0.5})^3R_b. \label{eqn:delta}
\end{equation}

In reality, $\delta$ may be comparable to the size of nano/micro-sized particles, resulting in two different modes regarding the collection of surface particles into the top jet drop. When $2r_p > \delta$ (Regime I, Figure \ref{fig:mechanism}(B)), a portion of the particles may be left behind in the bulk water due to the small jet drop volume \cite{blanchard1989ejection}, though the particle inertia can be neglected. While for $2r_p < \delta$ (Regime II, Figure \ref{fig:mechanism}(C)), a majority of the particles may follow the motion of the subsurface liquid shell and enter the top jet drop \cite{sakai1988enrichment}. Figure \ref{fig:modelprediction}(A) shows a map of these two regimes separated by the dash-dotted line ($2r_p = \delta$) with respect to $r_p$ and $R_b$. The size range of viruses ($r \approx 10-300$ nm) \cite{louten2016virus, colson2017mimivirus} and bacteria ($r \approx 0.1-5$ $\mu$m) \cite{levin2015small} are denoted by the blue and yellow regions, respectively. Bubbles with $R_b < 15$ $\mu$m and $R_b > 2.5$ mm in water cannot produce jet drops due to the inhibitions of viscous and gravity effects, respectively \cite{walls2015jet, veron2015}, and thus are not shown on the regime map. Since smaller bubbles are much more prevalent in natural bodies of water as compared to larger ones \cite{deane2002scale, blenkinsopp2010bubble}, the enrichment of microbes due to bubble bursting jetting may largely occur in regime I. We also summarize the parameter range of the prior studies \cite{sakai1988enrichment,blanchard1970,bezdek1972surface,hejkal1980water,blanchard1972concentration,blanchard1981bubble} and our experiments on the regime map. To our best knowledge, most existing studies (except for the experiments performed by Sakai et al. \cite{sakai1988enrichment}) as well as this study (including a set of data located at the boundary) are situated within regime I. Nonetheless, quantitative modeling of the enrichment effect for the jet drops in both regimes is still lacking, which will be the focus of the following discussion.

\subsection*{Scaling laws for the enrichment factor}

We first propose a theoretical argument to predict the enrichment factor $EF$ in regime I. Because the particle diameter $2r_p$ is larger than the thickness $\delta$ of the subsurface liquid shell forming the top jet drop, we assume that only a portion $\delta/(2r_p)$ of the particles captured by the bubble will be collected into the top jet drop, i.e., the particle number in the top jet drop $N_d = N_a\delta/(2r_p) = 3\pi R_bH\delta C_b/2$. Then, $EF$ can be obtained as
\begin{equation}
EF=\frac{3H}{8R_b}, \label{eqn:EF 1}
\end{equation}
which indicates that $EF$ is only determined by the normalized bubble rising distance $H/R_b$. Specifically, $EF$ increases with the bubble rising distance $H$ and decreases with the bubble radius $R_b$, while being independent of the particle radius $r_p$ in regime I. Such a scaling law captures the findings of our experiments well (Figures \ref{fig:raw data}(B) and \ref{fig:raw data}(C)). Furthermore, the $EF$s are collapsed to an approximately linear curve as predicted by Eq. \ref{eqn:EF 1} when normalizing $H$ with $R_b$ for different $R_b$ and $r_p$ over a wide parameter range, as shown in Figure \ref{fig:modelprediction}(B). We also compare our model with experimental data from prior studies \cite{blanchard1970,bezdek1972surface,hejkal1980water,blanchard1972concentration,blanchard1981bubble} in Figure \ref{fig:modelprediction}(C). We note that these previous experiments adopted a relatively complicated biological system to study the enrichment of active rod-shaped (serratia marcescens \cite{blanchard1970,blanchard1981bubble,blanchard1972concentration} or serratia marinorubra  \cite{bezdek1972surface,hejkal1980water}, 0.25 $\mu$m in radius and 1-2 $\mu$m in length) or spherical (micrococcus euryhalis \cite{hejkal1980water}, 0.25-1.5 $\mu$m in radius) bacteria. The deviations of reported $EF$ of these bacteria may be attributed to the different bacteria surface properties and metabolites of the bacteria (such as the secretion of biosurfactants which may affect the surface tension of water) caused by varying culture conditions \cite{burger1985droplet,syzdek1985influence} and age \cite{hejkal1980water}, as well as the active movement of the bacteria \cite{lushi2014fluid, Zhang2010Collective}. Nevertheless, the overall trend yielded by most of these prior studies follows the prediction of Eq. \ref{eqn:EF 1}. This comparison with previous work highlights the important role of bubble hydrodynamics in pathogen transmission induced by bubble bursting, and the potential of our model for the prediction of nano/micro-sized organism enrichment in bubble bursting jet drops.

In addition, some prior experimental studies reported that the $EF$ of bacteria decreases rapidly when $R_b < 0.2-0.3$ mm, approaching unity when $R_b \approx 0.1$ mm \cite{blanchard1970, bezdek1972surface}; there were no experimental reports with a smaller $R_b$ due to the limitations of bubble formation. A proposed explanation for this observation is that fewer bacteria are drawn into the jet drop as most bacteria are left behind in the bulk water when the bacteria size is larger than $\delta$ \cite{blanchard1983production, blanchard1989ejection}. However, our current work demonstrates a continuously increasing particulate $EF$ with decreasing $R_b$ within a wide range of $R_b$ (0.19-1.09 mm) in regime I with $2r_p > \delta$, showing $EF$ can be up to $571 \pm 86$ for $R_b=0.19$ mm attributed to the combination effect of bubble scavenge and bursting on the transport of nano/micro-sized particles. Thus, our study may help rule out the effect of bubble hydrodynamics as an explanation for the prior finding of a much smaller $EF$ of bacteria than expected at small $R_b$. We speculate that the previous experimental observation may originate from the statistics of bacteria sampling and culturing of the jet drops. In these experiments, the average bacteria count in the top jet drop generated by such small bubbles is much less than 1, which may be inadequate for the purpose of sampling \cite{bezdek1972surface, quinn1975breaking}. Additionally, the hydrodynamic stresses induced by jet breakup increase exponentially with decreasing jet drop (or bubble) radius \cite{mcrae2021aerosol}. It is likely that a portion of the bacteria are killed during these experiments with such small bubbles \cite{walls2017quantifying,hariadi2015determining}, preventing them from being cultured and thereby causing an underestimate in the original bacteria concentration in the jet drops.

For the particulate enrichment in regime II with $2r_p < \delta$, we assume all the particles captured by the rising bubble are concentrated in the top jet drop, i.e. $N_d = N_a$. Thus, we have
\begin{equation}
EF \sim \frac{423r_pH}{(1-(Oh/0.031)^{0.5})^3R_b^2}, \label{eqn:EF 2}
\end{equation}
which shows that $EF$ is mainly determined by both $H/R_b$ and the particle to bubble radius ratio $r_p/R_b$, considering the small change of $Oh$ ($= 0.003-0.008$) in our experiments. Different from that in regime I, $EF$ is influenced by particle radius $r_p$, and proportional to $R_b^{-2}$ in regime II. A comparison between the prediction of Eq. \ref{eqn:EF 2} and the experimental results of Sakai et al. \cite{sakai1988enrichment} is presented in Figure \ref{fig:modelprediction}(D). This comparison shows that $EF$ agrees well with Eq. \ref{eqn:EF 2} under different bulk particle concentrations, suggesting the validation of our model for the prediction of particulate enrichment in regime II.

It should be noted that the capture of particles by the rising bubble surface may be saturated under high rising distances $H$ and particle concentrations $C_b$, when the surface concentration of the particles reach a maximum value, and a maximum $EF$ may occur \cite{blanchard1981bubble,chingin2018enrichment}. In this study, we performed experiments at a parameter range without such a particle saturation to identify the effects of the bubble and particle dimensions, as well as the bubble rising distance. The maximum attainable particulate $EF$ in the jet drop and the effects of the related control parameters deserve more attention in future studies. In addition, smaller bubbles with a smaller $Re$ during rising may invalidate the potential flow assumption, necessitating a modification of particle-bubble collision model \cite{dai2000particle}. For particulates with $r_p < 0.1$ $\mu$m (such as surfactant molecules \cite{chingin2018enrichment}) or $r_p > 1$ $\mu$m (with a relative large $St$ number), the bubble scavenge may involve diffusive and inertia effects, respectively \cite{chingin2018enrichment, dai2000particle}. Furthermore, microbes in practical situations can exhibit different surface wettabilities or surface charges \cite{van1997determination, gannon1991relationship}, which may change the particle attachment as well as the surface mobility of bubbles \cite{walls2014moving}. Considering all these factors could extend our model to a wider parameter range and a more biorelevant dispersal environment in future work.  

\section*{Conclusion}

In summary, we investigate experimentally and theoretically the enrichment of nano/micro-sized particles in bubble bursting jet drops. Via high-speed visualization and direct measurement, we rationalize the detailed mechanism of particulate enrichment, resulting from a two-stage hydrodynamic process: the suspended particles are first intercepted by the rising bubble, attaching to the bubble surface before being skimmed off and concentrated into the jet drop originating from a thin liquid shell beneath the bubble surface during bubble bursting. By modeling the scavenge and collection of particles from the bulk fluid to the jet during bubble rising and bursting, we propose scaling laws for the particle enrichment factor in the top jet drop, which quantitatively agree with our experiments and prior studies \cite{sakai1988enrichment} on the enrichment of spherical particles. Furthermore, the proposed model reasonably captures the overall trend of enrichment factor of active bacteria with a wide range of experimental parameters reported in previous studies \cite{blanchard1970,bezdek1972surface,hejkal1980water,blanchard1972concentration,blanchard1981bubble}, thus characterizing the multi-phase fluid dynamics behind the water-to-air transmission of nano/micro-sized organisms via bubble bursting jets. 

Regarding the water-to-air disease transmission from toilets \cite{2021Abney}, swimming pools \cite{2005AngenentMolecular}, tubs \cite{1997EmbilPulmonary}, and wastewater plants \cite{2002BauerBacteria}, this study may provide potential guidelines to predict the pathogen load in bubble bursting aerosols and the infection risks resulting from their inhalation \cite{azimi2021mechanistic, lelieveld2020model}, advancing the evaluation and prevention of the indoor and community disease transmission related to bubble bursting \cite{bourouiba2020, 2021LouThe}. On a planetary scale, modeling the compositions of sea spray aerosol is essential since it dictates the radiant properties of the atmosphere, in addition to impacting atmospheric pollution, global climate, and ecological and material cycles \cite{deike2022mass, Brooks2018, hariadi2015determining, 2018schiffer, 1997Colin}, with nanoplastics and microplastics found in sea spray aerosols being of major concern in recent years \cite{masry2021experimental, allen2020examination}. By advancing the quantification of the particulate concentration in bubble bursting aerosols, our study may be incorporated into a global model for more accurate prediction of sea spray aerosol compositions.

\section{Materials and Methods}

\subsection{Materials}
Amino-modified polystyrene (PS) particles (spherical, fluorescent pink, 1.0 wt\% aqueous suspension) with a mean radius of $r_p$ = 0.25 $\mu$m were purchased from Spherotech, and amino-modified PS particles (spherical, fluorescent orange, 2.5 wt\% aqueous suspension) with $r_p$ = 0.5 and 1 $\mu$m were purchased from Sigma-Aldrich. These particle suspensions with $r_p$ = 0.25, 0.5 and 1 $\mu$m were diluted with DI water (resistivity = 18.2 M$\Omega$ cm, Smart2Pure 3 UV/UF, Thermo Fisher Scientific) by 64000, 20000, and 2500 times respectively, for a number concentration of $C_b \approx 2.3 \times 10^6$/mL used in the experiment. Experiments were performed at $22 \pm 2^\circ$C and 1 atm. The surface tensions of the aqueous particle suspensions were measured using the pendant-drop method as $\gamma=0.073$ N/m. In this study, $\rho_w$ = 998.2 kg/m$^3$, $\mu$ = 0.89 mPa s and $\rho_p$ = 1050 kg/m$^3$.

\subsection{Methods}
A square transparent acrylic tank with a cross-sectional area of $30 \times 30$ mm$^2$ was designed to hold the aqueous particle suspensions (Figure \ref{fig:experiments}(A)), which were agitated with a vortex mixer (Thermo Scientific) to ensure homogeneity prior to the conduct of experiments. A micropipette (with tip diameters of 1, 5, and 30 $\mu$m, from Fisher Scientific) or a stainless blunt steel needle (27 Gauge) was used to generate gas bubbles at a depth $H$ beneath the water surface, via a syringe pump (11 Pico Plus Elite, Harvard Apparatus) with a bubbling frequency of $<$ 6 bubbles per minute. The bubble formation in the bulk water (Figure \ref{fig:experiments}(B)) and bursting at the water surface (Figure \ref{fig:experiments}(C)) were visualized by a high-speed camera (FASTCAM Mini AX200, Photron) equipped with focus lens of magnification $3-6\times$, illuminated by an LED panel. A frame rate of 6400-20000 frames per second and an exposure time of 5-50 $\mu$s were used. The obtained images were further processed using Fiji (ImageJ) \cite{schindelin2012fiji} to calculate the bubble radius $R_b$ and the top jet drop radius $r_d$, which are confirmed to remain constant during every experiment. A glass slide is placed above the water surface, with the distance between the slide and the water surface being adjusted to ensure that only the top jet drop is collected by the slide, which is confirmed by the high-speed footage (Figure \ref{fig:experiments}(D)). The collected drop is dried before being examined under a microscope (Nikon ECLIPSE Ti2, Nikon) with an objective lens of 10X, 20X or 40X, and the the number of particles in the entire jet drop ($N_d$) is determined via an in-house image analysis code using MATLAB 2021a (Figure \ref{fig:experiments}(E)). Similarly, a pipette is used to collect a drop from the bulk suspension, with the drop volume ($V_b$) being verified via the high-speed camera before being dried on a glass slide and examined under the microscope (Figure \ref{fig:experiments}(F)), the same MATLAB image analysis code is used to determine the number of particles in the entire bulk drop ($N_b$). In every experiment, at least 10 top jet drops and 3 bulk drops are examined to determine the average $N_d$ and $N_b$, respectively.

\begin{acknowledgement}

This work is partially supported by American Chemical Society Petroleum Research Fund Grant No. 61574-DNI9 (to J.F.)

\end{acknowledgement}

\begin{suppinfo}

The Supporting Information is available free of charge on the
ACS Publications website at XXX

Video of bubble formation from a micropipette (AVI)

Video of bubble bursting at the surface of particle suspension (AVI)

Video of the collection of the top jet drop by a glass slide (AVI)

\end{suppinfo}

\bibliographystyle{}
\bibliography{achemso-demo}

\end{document}